\documentclass{article}

\begin{document}

\title {Magnetic and Electric Fields around the Black Hole in Cyg X-1}

\author {$Gnedin\, Yu.N.^{(1)}, Borisov\, N.V.^{(2)},
Natsvlishvili\, T.M.^{(1)},$\\$Piotrovich\, M.Yu.^{(1)},
Silant'ev\, N.A.^{(1,3)}$}

\maketitle

\begin{center}
{\small (1) Central Astronomical Observatory at Pulkovo,
Saint-Petersburg, Russia.

(2) Special Astrophysical Observatory, Nizhnii Arhyz, Russia.}

(3) Instituto National de Astrofisica, Optica y Electronica,
Puebla, Mexico.
\end{center}

\begin{abstract}
Analysis of polarimetric observations of X-ray binary Cyg X-1/HDE
226868 including the data obtained by BTA-6m allows to estimate
the magnetic field magnitude near the inner radius of the
accretion disk. The magnetic field magnitude occurred to be $\sim
10^{8}$ G. For power law of radial dependence of magnetic field
into an accretion disk we estimates the value of an index of power
law. For the Cyg X-1/HDE 226868 system the value of this index
appears non less then two. If one accepts as a characteristic
scale of a magnetic field generation region the dyadosphere
radius, one can estimate the charge magnitude of a black hole. For
Cyg X-1 this magnitude appears to be $\sim 0.01M\sqrt{G}$, where
$M$ is a black hole mass.
\end{abstract}

\section{Introduction}

Cygnus X-1 is a X-ray binary star system consisting of a compact
object of at least 8 solar masses and of an O9.7 Iab supergiant
with the effective temperature $T_{e}=30000$K and luminosity
$L\approx 10^{38} erg\, s^{-1}$. This magnitude is approximately
ten times higher than the X-ray luminosity of the compact object.
X-ray and optical fluxes of this system are varied with the
orbital period $P_{orb}=5^{d}.6$. The light curve determines the
mass function that appears quite low $f(M)=0.25M_{\odot}$.
Nevertheless, taking into account the value of the inclination
angle and the typical magnitude of OB-supergiant mass one can
estimate the compact object mass not less than $8M_{\odot}$. It
means that the X-ray compact object in the Cyg X-1 system is the
black hole (see the last review by Cherepaschuk, 2001).

The observed X-ray spectrum of Cygnus X-1 is well described as a
radiation from a geometrically thin plasma accretion disk around
the black hole. The theory of such disks is now well developed,
the basic classical work having made by Shakura and Sunayev
(1973). The disk model reproduces quite well as the soft so hard
spectral components in high or low states of a disk. The main
difficulty of the modern models is to explain the unusual
transitions between these states.

There is an exciting problem of existence of magnetic field in the
nearest environment of a black hole. This problem appears tightly
connecting with another key problem how to extract energy from a
rotating black hole. Many authors have considered various
alternative mechanisms for extracting energy from a rotating black
hole. Among them the most promising one is the well-known
Blandford-Znajek mechanism (Blandford and Znajek, 1977). In this
mechanism a Kerr black hole is assumed to connect with surrounding
matter with magnetic field lines. The magnetic field lines thread
the black hole's horizon and its rotation twists the magnetic
field lines and transports energy and angular momentum from the
black hole to the accretion disk (Blandford, 2001; Li, 2002; Li
and Paczynski, 2000 and references therein).

A magnetic field connecting a black hole to a disk has important
effects on the balance and transfer of energy and angular
momentum. The global structure of black hole magnetosphere
involving axisymmetric magnetic field and plasma injected from an
accretion disk has been extensively investigated for explaining
various observational features of transient X-ray binaries and
AGNs (see,e.g., for review by Beskin, 1997). Recently Robertson
and Leiter (2002) claimed an evidence for intrinsic magnetic
moments in galactic black hole candidates.

Another new idea was presented recently by Ruffini et al.(2001,
2002) and Punsly (2001). It has been suggested that the observed
features of gamma-ray bursters can be modelled by the presence of
an electrically charged black hole. Preparata et al. (1998 and
2002) present a model of GRB 971214. They claimed that basic
energy requirements of GRB sources can be easily accounted for by
a pair creation process occurring in, so-called, the "Dyadosphere"
that can be considered as the ergosphere of a charged black hole.
The "Dyadosphere" is defined by Preparata et al.(2002) as the
region outside the horizon of an electromagnetic black hole (EMBH)
where electromagnetic field exceeds the critical value for
$e^{+}e^{-}$ pair production.

The energy of a fast rotating BH can be transferred into energy of
strong relativistic jets. Namely this process was considered by
Blandford and Znajek (1977). They estimated the energy value being
extracted from a fast rotating BH due to a magnetic field strength
$B$:
\begin{equation}
L_{x}\sim 6\times 10^{38} a \left(\frac{M_{BH}}{M_{\odot}}
\right)^{2}\left(\frac{B}{10^{8}G}\right) erg\, s^{-1}
\end{equation}
where $a=\frac{l}{M}$ is the specific angular momentum of a black
hole.

However the strong evidence of existence of the global magnetic
field in the nearest environment of BH in the close binary systems
was obtained from the direct polarimetric observations. Michalsky
et al.(1975) have presented an evidence of variable circular
polarization of the binary system Cyg X-1/HDE 226868, that was
varied with the period of the binary system $P_{orb}=5^{d}.6$.
They claimed the magnetic origin of this polarization. Below we
shall consider all combined observed data that give possibility to
determine the magnetic field strength near the black hole in the
Cyg X-1/HDE 226868 binary system. The results of
spectropolarimetric observations in the spectral line HeII
$\lambda 4686$ made with BTA-6m telescope will be also presented.
These data allow to derive the radial distribution of magnetic
field in the accretion disk. If the characteristic size of the
magnetic field region coincides with the Dyadosphere radius one
can determine the value $Q$ of charge of black hole in the Cyg
X-1/HDE 226868 binary system.

\section{Polarimetric Observations of Cyg X-1 and Determination of
the Magnetic Field Strength near the Black Hole}

The first determination the magnetic field strength near the
compact object in the Cyg X-1/HDE 226868 binary system have been
made in a result of observing circular optical polarization of
this system (Kemp et al., 1972, Michalsky et al., 1975, 1977). The
magnitude of the circular polarization have been estimated at the
level of $P_{V}=(4.8\pm 0.5)\times 10^{-4}$. The detail analysis
shows that this polarization can not be of interstellar origin,
i.e., can't be generated, for instance, in a result of conversion
of the intrinsic linear polarization of this system into the
circular one onto the oriented interstellar dust grains (see, for
example, Dolginov et al., 1995), because this mechanism requires
too high value compare to observational intrinsic linear
polarization of this system. It requires the magnitude of the
intrinsic linear polarization, at least, at the level
$P_{l}\approx 2\%$ that contradicts to observations. The real
polarimetric observations give the magnitude of intrinsic linear
polarization at the level only $P_{l}\approx 0.2\%$. Therefore
seems naturally the explanation of observed circular polarization
as a magnetic origin. This polarization can appear in a result of
scattering of a light by electrons in surrounding plasma (see, for
example, Dolginov et al., 1995, Gnedin and Silant'ev, 1997). The
magnitude of magnetic field is readily estimated by
\begin{equation}
P_{V}\sim \frac{\omega_{B}}{\omega}\sim 6\times
10^{-2}\left(\frac{B}{10^{8}G}\right)\left(\frac{\lambda}{4500\textmd{\AA}}\right)
\end{equation}
where $\omega_{B}=\frac{eB}{m_{e}c}$ is the cyclotron frequency,
and $\omega$ is the radiation frequency. The magnetic field
strength $B\sim 10^{6}$G corresponds namely to the observed value
$P_{V}\approx 5\times 10^{-4}$ circular polarization. It is
evident that the magnetic field of such magnitude cannot exist at
the surface of the supergiant HDE 226868. Therefore one can make
the conclusion that an accretion disk namely is the origin of the
magnetic field of such magnitude.

One ought to consider the value $B\approx 10^{6}$G as only the
lower limit of the real magnetic field value in the accretion disk
because of the dilution effect by the stellar optical light of HDE
226868 itself. Usually one estimates the accretion disk optical
luminosity at the level of $\geq 1\%$ of the total optical
luminosity of the binary system Cyg X-1/HDE 226868. It means that
the real strength of magnetic field is of order of $B\sim
10^{7}\div 10^{8}$G.

The next important stage of searching magnetic field in the system
Cyg X-1/HDE 226868 is connected with the first polarimetric
observations of this system in X-ray spectral range. The X-ray
polarimetric observations have been made by Long et al.(1980) with
the Bragg crystal polarimeter aboard OSO8. The marginal detection
of the time-averaged polarization was given for Cyg X-1 at the
level: $P_{l}(2.6KeV)=2.4\%\pm 1.1\%$ and $P_{l}(5.2KeV)=5.3\%\pm
2.5\%$. Unfortunately, the observational results occurred at the
low confidence level, only $\sim 2\sigma$. Nevertheless, if one
suggest that the decrease of net polarization for the energy
$E=2.6KeV$ due to the effect of Faraday depolarization (see Gnedin
and Silant'ev, 1980, 1984, 1997, and Dolginov et al., 1995) it is
possible to derive the magnetic field strength in the nearest
environment of the black hole where X-ray radiation is generated.

The angle of Faraday rotation $\chi$ is determined by the
expression (Gnedin and Silant'ev, 1980):
\begin{equation}
\chi=\frac{1}{2}\delta\tau_{T}\cos{\theta}; \quad
\delta=\frac{3\omega_{B}c}{2r_{e}\omega^{2}}\cong
1.2\left(\frac{B}{10^{6}G}\right)\left(\frac{1KeV}{\hbar\omega}\right)^{2}
\end{equation}
where $\tau_{T}$ is the optical thickness of the region of
electron scattering, $\theta$ is the angle between directions of
magnetic field $\overrightarrow{B}$ and radiation propagation
$\overrightarrow{n}$, $r_{e}=\frac{e^{2}}{m_{e}c^{2}}$ is the
classical electron radius.

If the magnetic field is increased the angle $\chi$ increases too
and partly polarized scattered radiation begins to undergo Faraday
rotation. The rotation angles $\chi$ are different for photons
scattered in different volumes alone the line of sight
$\overrightarrow{n}$, and in a result the total radiation from all
volume elements will be depolarized (see Fig.10 from the review by
Gnedin and Silant'ev, 1997).

The spectra of polarized radiation scattered in a spherically
symmetric magnetized envelope, and also in optically thick
scattering disk have been calculated by Dolginov et al., 1995,
Gnedin and Silant'ev, 1997, Silant'ev, 2001.

For an accretion disk the Stokes parameters have been derived by
Silant'ev, 2001:
\begin{eqnarray}
Q(\overrightarrow{n},\overrightarrow{B})=-\frac{F}{2\pi
J_{1}}\frac{1-g}{1+g}\frac{(1-\mu^{2})(1-k\mu)}{(1-k\mu)^{2}+(1-q)^{2}\delta^{2}\cos{\theta}^{2}}\nonumber
\\ U(\overrightarrow{n},\overrightarrow{B})=-\frac{F}{2\pi
J_{1}}\frac{1-g}{1+g}\frac{(1-\mu^2)(1-q)\delta\cos{\theta}}{(1-k\mu)^{2}+(1-q)^{2}\delta^{2}\cos{\theta}^{2}}
\end{eqnarray}
Here $\overrightarrow{n}$ is the line of sight direction, $\theta$
is the angle between the line of sight and the magnetic field
$\overrightarrow{B}$, $\mu=\cos{\vartheta}$, where $\vartheta$ is
the angle between the line of sight and the normal to the surface
of the accretion disk, $q=\frac{\sigma_{a}}{\sigma_{T}}$ is the
ratio between the cross-sections of absorption and electron
scattering. The values of constants $J$, $g$ and $K$ are tabulated
by Silant'ev, 2001. He also published there the dependencies of
polarization of an accretion disk radiation on various magnitudes
of $\mu$, $q$ and $\delta$.

From Eq.(4) it is evident that with increase of the depolarization
parameter $\delta$ and the magnetic field strength $B$
respectively, the polarization drops as $P_{l}\sim
\frac{1}{\delta}\sim \frac{1}{\lambda^{2}B}$ where $\lambda$ is
the radiation wavelength. For the quite large values of $\delta\gg
1$ the angular dependence of net polarization on the angle
$\theta$ has narrow maximum into the angular interval
$\Delta\theta\sim \frac{1}{(1-q)\delta}$.

Using the polarimetric observations in X-ray spectral range one
can estimate the magnetic field strength near the black hole
itself, if one suggests that the decrease of polarization at the
energy $E=2.6KeV$ is due to the depolarization effect. Then the
requirements of $\delta (E=2.6KeV)\gg 1$ allows to get the
following estimation of the magnetic field strength in the nearest
environment around the black hole in Cyg X-1 system: $B\geq
3\times 10^{7}G$.

Independent estimation of a magnetic field of optical radiation
region around the Cyg X-1 black hole can be too made via the
measured intrinsic linear polarization of Cyg X-1/HDE 226868. This
polarization has been discovered by Nolt et al. (1975). The
amplitude of the variable polarization occurred quite low at the
level of $\sim 0.2\%$, with the very complex pattern of
variability. Nolt et al (1975) claimed the discovery of various
types of polarization variability with $39^{d}$ and $78^{d}$
periods and of long-term variability. The possible mechanisms of
this variability have been in detail discussed by Karitskaya,
1979, 1981 and by Bochkarev et al., 1979. Unfortunately, the real
explanation of observed linear polarization of Cyg X-1 is absent
up to date.

If one accepts the accretion matter around the black hole as a
real source of the observed optical polarization it is necessary
to increase the intrinsic optical polarization of the black hole
environment at least in $\sim 100$ times, because of the high
dilution of polarized radiation by the optical light of the
supergiant HDE 226868. Its luminosity is estimated at the level of
$L_{0}\approx (1\div 3)\times 10^{39} erg/s$. Let us remind the
X-ray luminosity of Cyg X-1 is estimated as $L_{X}\sim 8\times
10^{37} erg/s$ (see, for example, Cherepaschuk, 2001). We consider
the case when the optical radiation is generated near the black
hole as a result of reprocessing of X-ray by nearest accretion
matter. It means that the real magnitude of net linear
polarization from the nearest accreted mater can reach magnitude
at the level $P_{l}\geq 10\%$. Such high value polarization can be
dare say produced in a result of single scattering in outflows
such as a magnetized wind or dynamical corona.

One can estimate the resulting polarization by using calculations
of polarization of radiation from a central light source (a star)
surrounded by a thin magnetized shell made by Dolginov et al,
1995, (see the paragraph 9 of section 4 of their book). In a rough
model of magnetized spherically symmetric electron shell
surrounding the accretion disk of the central black hole the net
polarization can reach magnitude $P_{l}\sim 10\%$ for
$\delta\tau_{Sh}\approx 10$ where $\tau_{Sh}$ is the optical
thickness of a shell respect to electron scattering. In true it is
valid only for the case when the line of sight is perpendicular to
the magnetic field direction. If one suggests that
$\tau_{Sh}\approx 0.1$, then the magnetic field strength in
optical region of the disk can be easily estimated from the
expression
\begin{equation}
\delta = 0.8
\left(\frac{B}{1G}\right)\left(\frac{\lambda}{1mM}\right)^{2}
\end{equation}
The Eq.(5) allows to estimate the magnetic field strength in the
region of optical radiation generation as $B\approx 500G$. A
number of models of accretion disk gives the value of the corona
inner radius as $\sim 100 R_{g}$, where
$R_{g}=\frac{2GM_{BH}}{c^{2}}$ is the gravitational radius. For
dipolar magnetic field the estimation of the magnetic field
strength in nearest environment of the black hole gives
$B(3R_{g})\sim 10^{7}G$ that is close to previous estimations.

\section{Estimation of a Charge of the Black Hole in Cyg X-1/HDE
226868 Binary System}

Accreting black holes can release enormous amounts of energy to
their surroundings in quite different forms. The idea of a black
hole endowed with electromagnetic structure (EMBH) has recently
become very popular. Still in 1971 Ruffini and Wheeler proposed
the famous "uniqueness theorem" stating that black holes can only
be characterized by their mass-energy $E$, charge $Q$ and angular
moment $l$. This idea has been recently developed by Ruffini
(2002), Ruffini and Vitagliano (2002) (see refs therein). Various
models of particle accelerations near black holes have been widely
discussed at last time in connection with the studies of
synchrotron radiation and inverse Compton scattering from jets
observed over a wide spectral range, from radio to gamma rays.
Preparata et al.(2002) formulated the derivation of, so-called,
"Dyadosphere" that was defined as the region outside the horizon
of an EMBH where the electromagnetic field exceeds the critical
value for $e^{+}e^{-}$ pair production. They expressed the outer
radius of dyadosphere in the form:
\begin{equation}
R_{ds}=\left(\frac{\hbar}{m_{e}c}\right)^{\frac{1}{2}}
\left(\frac{GM}{c^{2}}\right)^{\frac{1}{2}}
\left(\frac{m_{p}}{m_{e}}\right)^{\frac{1}{2}}
\left(\frac{e}{q_{p}}\right)^{\frac{1}{2}}
\left(\frac{Q}{\sqrt{G}M}\right)^{\frac{1}{2}} \approx 1.12\times
10^{8}\sqrt{\mu\xi}\, cm
\end{equation}
where $m_{p}=(\hbar c / G)^{0.5}$ is the Planck mass and
$q_{p}=(\hbar c)^{0.5}$ is the Planck charge, $\mu
=\frac{M}{M_{\odot}}$, $\xi =\frac{Q}{Q_{max}}$

Another model of EMBH has been recently developed by Shatsky
(2001, 2003, Shatsky and Kardashev, 2002). He suggested a
mechanism that combines a unipolar inductor and strong
gravitational black hole effects. In the result he showed that the
black hole electric charge can be expressed in terms of the
magnetic field at the disk center $B_{0}$ as follows:
\begin{equation}
Q=\frac{\pi}{2}\Omega R^{2} a B_{0}
\end{equation}
where $\Omega$ is the Keplerian rotation frequency, $R$ is the
disk radius and $a$ is the distance from the center where located
two charges $\pm Q$. In this model the electric field of the disk
element can be presented as the field from two charges $\pm Q$,
located inside the disk.

For estimation of a charge magnitude of the Cyg X-1 black hole we
use the value of a intrinsic magnetic moment of this black hole
that has been recently estimated by Robertson and Leiter (2002).
They tested the hypothesis that the power law part of the
quiescent emissions of the black hole candidates might be
magnetospheric origin. They derived proposed magnetic moments and
rates of spin for them and predicted their quiescent luminosities.
For Cyg X-1 the estimated magnitude of the intrinsic magnetic
moment appeared $m =1.26\times 10^{30}\, G\, cm^{3}$. Let us
suggest that the characteristic scale dimension of magnetic field
generation and the dyadosphere radius coincide one to other. Then,
using the estimated by Robertson and Leiter value of the intrinsic
magnetic moment of the black hole in Cyg X-1 and obtained from
polarimetric observations the magnetic field strength $B\approx
3\times 10^{7}\,G$ one may obtain the value of the Cyg X-1
dyadosphere radius directly from polarimetric observations:
\begin{equation}
R_{ds}(Cyg X-1)=3.5\times 10^{7}\, cm
\end{equation}
Comparing Eqs. (6) and (7) one can estimate the electric charge
magnitude of the black hole in Cyg X-1:
\begin{equation}
Q=7.8\times 10^{-3}\left(\frac{10M_{\odot}}{M_{BH}}\right)Q_{max}
\end{equation}

Quite similar estimation can be obtained if one proposes the
electron-position jet region radius in the Cyg X-1 nonthermal
corona as the magnetic field region radius. Maccarone and Coppi
(2002) estimate this radius as $R_{car}\approx 10^{8}\, cm$. Then
the magnitude of a probable electric charge for the Cyg X-1 black
hole appears as
\begin{equation}
Q\approx 0.08\left(\frac{10M_{\odot}}{M_{BH}}\right)Q_{max}
\end{equation}

Now we can estimate the rotation rate of the black hole in Cyg X-1
with use of Shatsky model (see Eq.(7)). Accepting the distance
$a\approx h\approx (0.1\div 1)3R_{g}$, where $h$ is the height of
the accretion disk, one can estimate the rotation rate $\Omega$ of
the black hole that is producing the charge $Q\cong 0.08Q_{max}$:
\begin{equation}
\Omega\approx \frac{2Q}{\pi a R^{2} B_{0}}\approx 20
\left(\frac{Q}{0.1}\right)\left(\frac{3\times
10^{7}G}{B_{0}}\right)
\end{equation}
It means that rotation period of the Cyg X-1 black hole $P\approx
0^{s}.03\div 0^{s}.3$ depending on our estimation of the charge
magnitude (see Eqs. (9), (10)).

\section{The Radial Structure of Magnetic Field in the Accretion Disk}

The spectral and spectropolarimetric observations of the Cyg
X-1/HDE 226868 were made by BTA-6m telescope at 2001 July, with
UAGS spectrograph and CCD analyzer of circular and linear
polarization. The diffraction slit with a dispersion $1300\,
mm^{-1}$ have been used. This slit provides the spectral
resolution $1.5\textmd{\AA}/pixel$ in the $3500-9000\textmd{\AA}$
range. During observations the seeing was not higher $1".5$. The
reprocessing was made with the standard MIDAS program. The results
of observations are presented at Fig.1. The spectropolarimetric
observations have been carried out near the spectral line of $HeII
\lambda 4686$, that is usually proposed to be generated in an
accretion disk. The upper limit of the circular polarization net
in $HeII\lambda 4686$ line appears as $P_{v}<0.2\%$. One can
estimate the constraints of a magnetic field strength in the
region of $HeII$ lines generation: $B(R_{HeII})<10^{3}\,G$.

The $HeII\lambda 4686$ line is produced in the region of the high
temperature $T_{e}(R_{HeII})\approx 10^{5}\, K$. The standard
theory of an accretion disk (Shakura and Sunayev, 1973) determines
the following radial temperature distribution:
\begin{equation}
T_{e}(R)=T_{in}\left(\frac{R_{in}}{R}\right)^{3/4}
\end{equation}
where $R_{in}$ and $T_{in}$ are the radius and temperature at the
disk inner radius, respectively.

Suggesting that the inner radius corresponds to the marginal orbit
radius $R_{in}=3R_{g}=6GM/c^{2}$ and $kT_{e}\approx 1KeV$, i.e.
corresponds to X-ray temperature one can readily derive the helium
ionization radius
\begin{equation}
R_{HeII}=R_{in}\left(\frac{T_{in}}{T_{HeII}}\right)^{4/3}
\end{equation}
For the Cyg X-1 black hole with the mass magnitude
$M_{BH}=10M_{\odot}$ we have $R_{HeII}=1.4\times 10^{3}R_{g}$. If
one now suggests the power low of magnetic field radial
distribution
\begin{equation}
B=B_{0}\left(\frac{R}{3R_{g}}\right)^{-\delta}
\end{equation}
where $B_{0}=B(3R_{g})$, it is easy to determine the value
$\delta$ of a power law index. Choosing the values of magnetic
field strengths as $B_{0}=10^{8}\, G$ and $B(R_{HeII})\approx
10^{2}\, G$, one can derive the value $\delta\approx 2$. This is
the first experimental determination of the index of a magnetic
field radial dependence in accretion disks.

\section{Conclusions}

Polarimetric observational data of X-ray binary Cyg X-1/HDE 226868
allows to estimate the magnetic field strength near the last
marginal orbit of a black hole as $B_{0}\approx 10^{8}\, G$. If
one suggests that the characteristic scale size of the magnetic
field region is the same as a dyadosphere radius (Preparata et
al., 2002; Ruffini et al., 2003) one can estimate the charge value
of the Cyg X-1 black hole. It occurs at the level $\sim 1\%$ of
its maximal magnitude $M\sqrt{G}$. At the basis of BTA-6m
spectropolarimetric data of Cyg X-1 the radial distribution of the
magnetic field in the accretion disk was estimated. The power law
index appeared $\delta \geq 2$.

We are grateful to D.I. Makarov (SAO RAN) for the help in
obtaining of observational material.

This work was supported by the Russian Programs GNTP "Astronomy"
and by the Russian Foundation for Basic Research (Project
3-02-17223-a).


\begin{thebibliography}{99}
\bibitem{1}Beskin V.S., 1997, Usp. Fiz. Nauk, v.167, p.689.
\bibitem{2}Blandford R.D., 2001, Galaxies and their Constituents at Highest Angular Resolution., Proc. IAU Symp. 205, ed. R.T. Schilizzi (San Francisco, ASP 2001), p.10; astro-ph/0110397.
\bibitem{3}Blandford R.D., Znajek R.L., 1977, MNRAS, v.176, p.465.
\bibitem{4}Bochkarev N.G., Karitskaya E.A., Sunayev R.A., Shakura N.I., 1979, Sov.Astron.Zh., v.5, p.185.
\bibitem{5}Cherepaschuk A.M., 2001, Usp. Fiz. Nauk, v.171, p.864.
\bibitem{6}Dolginov A.Z., Gnedin Yu.N., Silant'ev N.A., 1995, in "Propagation and Polarization of Radiation in Cosmic Media", Gordon and Breach Publs., Amsterdam.
\bibitem{7}Gnedin Yu.N., Silant'ev N.A., 1980, SvAL, v.6, p.344.
\bibitem{8}Gnedin Yu.N., Silant'ev N.A., 1984, Ap.Sp.Sci., v.102, p.175.
\bibitem{9}Gnedin Yu.N., Silant'ev N.A., 1997, Basic Mechanisms in Light Polarization in Cosmic Media, Harwood Academic Publ., Amsterdam.
\bibitem{10}Karitskaya E.A., 1979, Sov.Astron.Circ., No 1088, p.1.
\bibitem{11}Karitskaya E.A., 1981, Sov.Astron.Zh., v.58, p.146.
\bibitem{12}Kemp J.C., Wolstencroft R.D., Swedlund L.B., 1972, Ap.J.Lett., v.173, L118.
\bibitem{13}Li L.-X., 2002, astro-ph/0202361.
\bibitem{14}Li L.-X., Paczynski B., 2000, Ap.J., v.534, L197.
\bibitem{15}Long K.S., Chanan G.A., Novick R., 1980, Ap.J., v.238, p.710.
\bibitem{16}Maccarone T.J., Coppi P.S., 2002, astro-ph/0204235.
\bibitem{17}Michalsky J.J., Stokes G.M., Stokes R.A., 1977, Ap.J.Lett., v.216, L35.
\bibitem{18}Michalsky J.J., Swedlund J.B., Stokes R.A., 1975, Ap.J.Lett., v.198, L101.
\bibitem{19}Nolt I.G., Kemp J.G., Rudy R.J., Rodostitz J.V., Caroff L.J., 1975, Ap.J.Lett., v.199, L27.
\bibitem{20}Preparata G., Ruffini R., Xue S.-S., 1998, Astron.Astrophys., v.338, L87.
\bibitem{21}Preparata G., Ruffini R., Xue S.-S., 2002, astro-ph/0204080.
\bibitem{22}Punsly B., 2001, Black Hole Gravitohydromagnetics, Springer.
\bibitem{23}Robertson S.L., Leiter D.J., 2002, Ap.J., v.565, p.447; 2002, astro-ph/0208333.
\bibitem{24}Ruffini R., 2002, astro-ph/0209264.
\bibitem{25}Ruffini R., Bianco C.L., Chardonnet P., Fraschetti F., Xue S.-S., 2001, Ap.J.Lett., v.555, L107, L113, L117; 2001, Nuovo Cim., v.116B, p.99.
\bibitem{26}Ruffini R., Vitagliano L., 2002, astro-ph/0209072.
\bibitem{27}Ruffini R., Wheeler J.A., 1971, Relativistic Cosmology from Space Platforms, in Proc. Conf on Space Phys, eds. Hardy V. and Moore H., E.S.R.O., Paris.
\bibitem{28}Shakura N.I., Synayev R.A., 1973, Astron. Astrophys., V.24, p.337.
\bibitem{29}Shatsky A.A., 2001, Zh.Eksp.Teor.Fiz., v.93, p.920 (gr-qc/0202068).
\bibitem{30}Shatsky A.A., 2003, astro-ph/0301535.
\bibitem{31}Shatsky A.A., Kardashev N.S., 2002, Astron.Zh., v.46, p.639, (astro-ph/0209465).
\bibitem{32}Silant'ev N.A., 2002, Astron.Astrophys., v.283, p.326.
\end{thebibliography}
\end{document}